\documentclass{aa}
\usepackage{graphicx,times}

\begin{document}

\title{X-ray/UV campaign on the Mrk~279 outflow:
density diagnostics in Active Galactic Nuclei
using \ion{O}{v} K-shell absorption lines}

\author{ J.S. Kaastra \inst{1}
         \and
         A.J.J. Raassen \inst{1,}\inst{2}
         \and
         R. Mewe \inst{1}\thanks{Deceased 4 May 2004}
         \and
	 N. Arav \inst{3}
	 \and
	 E. Behar \inst{4}
	 \and
	 J.R. Gabel \inst{3}
	 \and
	 G.A. Kriss \inst{5}
	 \and
	 D. Proga \inst{3}
	 \and
	 K.C. Steenbrugge \inst{1}
	          }
  
\offprints{J.S. Kaastra}
\mail{J.Kaastra@sron.nl}

\institute{ SRON National Institute for Space Research
              Sorbonnelaan 2, 3584 CA Utrecht, The Nether\-lands 
              \and
              Astronomical Institute "Anton Pannekoek", Kruislaan 403,
              1098 SJ Amsterdam, The Netherlands
              \and
	      CASA, University of Colorado, 389 UCB, Boulder, CO 80309-0389, USA
	      \and
	      Department of Physics, 
	      Technion-Israel Institute of Technology, Haifa 32000, Israel
	      \and
	      Space Telescope Science Institute, 3700 San Martin Drive, 
	      Baltimore, MD 21218, USA
              }

\date{Received  / Accepted  }

\abstract{
One of the main problems in modeling the ionised outflows in Active Galactic
Nuclei is the unknown distance of the outflowing wind to the central source.
Only if the density is known this distance can be determined through the
ionisation parameter. Here we study density diagnostics based upon \ion{O}{v}
transitions. \ion{O}{v} is known to have metastable levels that are density
dependent. We study the population of those levels under photoionisation
equilibrium conditions and determine for which parameter range they can have a
significant population. We find that resonance line trapping plays an important
role in reducing the critical densities above which the metastable population
becomes important. We investigate the K-shell absorption lines from these
metastable levels. Provided that there is a sufficient population of the
metastable levels, the corresponding K-shell absorption lines are detectable
and are well separated from the main absorption line originating from the
ground state. We then present the Chandra LETGS spectrum of the Seyfert 1
galaxy Mrk~279 that may show for the first time the presence of these
metastable level absorption lines. A firm identification is not yet possible
due to both uncertainties in the observed wavelength of the strongest line as
well as uncertainties in the predicted wavelength. If the line is indeed due to
absorption from \ion{O}{v}, then we deduce a distance to the central source of
one light week to a few light months, depending upon the importance of
additional heating processes.
}

\titlerunning{Density diagnostics in Active Galactic Nuclei
using \ion{O}{v} K-shell absorption lines}
\authorrunning{J.S. Kaastra et al.}

\maketitle

\section{Introduction}

Active Galactic Nuclei (AGN) have X-ray spectra that are dominated by
nonthermal emission from the immediate surroundings of a supermassive black
hole. Seyfert galaxies constitute the lower luminosity class of AGN. In a
significant fraction of all Seyfert galaxies the spectral signatures of the
so-called warm absorber are visible (e.g. Reynolds \cite{reynolds97}; George et
al.\ \cite{george98}). This warm absorber is very likely a photoionised outflow
from the accretion disk, with strong UV (for example Crenshaw et al.\
\cite{crenshaw99}) and X-ray absorption lines (Kaastra et al.\
\cite{kaastra00}). X-ray observations with high sensitivity show the presence
of absorption lines with a broad range of ionisation parameter $\xi = L/nr^2$,
where $L$ is the 1--1000~Ryd luminosity, $n$ the density and $r$ the distance
to the central source. A good example of this broad range in ionisation
parameter is the XMM-Newton Reflection Grating Spectrometer (RGS) spectrum of
\object{NGC~5548}, which shows the presence of K-shell absorption lines from
all oxygen ions between \ion{O}{iii}--\ion{O}{viii} (Steenbrugge et al.\
\cite{steenbrugge03}).

One of the main problems in modeling this outflow is that its distance $r$ from
the nucleus is hard to measure. Photoionisation modeling only yields the
product $nr^2$, as well as the column density $nd$ with $d$ the thickness of
the absorbing layer in the line of sight. Lacking a density, both the thickness
$d$ and the distance $r$ are essentially unknown. Gabel et al. (\cite{gabel03})
detected UV absorption lines from the metastable 2s\,2p~$^3$P triplet in
\ion{C}{iii} around 1175~\AA\ in the highest velocity outflow component in
\object{NGC~3783}. These measurements indicate a relatively high density of
order $10^{15}$~m$^{-3}$ corresponding to an upper limit of the distance to the
nucleus of 0.3~pc. However these density estimates are somewhat uncertain due
to the contamination of the \ion{C}{iii} resonance line at 977~\AA\ with
Galactic absorption in that source. Behar et al.\ \cite{behar03} pointed out
that the 2s\,2p triplet of \ion{C}{iii} can in fact be populated at low
densities. Only one fine-structure level in the triplet requires high densities
(see, e.g., Fig.~1 in Bhatia \& Kastner \cite{bhatia93}). Indeed updated
calculations by Gabel et al. (\cite{gabel04a}) show a lower density of $3\times
10^{10}$~m$^{-3}$ and a distance of 25~pc.

Other constraints on density hence distance may be derived from reverberation
studies of the warm absorber in response to continuum variations, by measuring
the recombination time scale. Long grating observations of NGC~3783 with 
XMM-Newton (Behar et al.\ \cite{behar03}) and with Chandra (Netzer et al.
\cite{netzer03}) have allowed for attempts to study the response of the
absorber to the ionising continuum through reverberation. However, since no
response was detected these could provide only lower limits (of the order of
0.5~pc) to the distance of the absorber from the central source. Reeves et al.
\cite{reeves04} did claim to see variability in the Fe-K absorption of that
source, albeit with CCD spectra. They deduce an upper limit to the distance
of the absorber to the central source of 0.02~pc.

In stellar coronae, with collisional ionisation equilibrium, several X-ray
emission lines are density-dependent, for example the forbidden and
intercombination lines of the \ion{O}{vii} triplet. As these lines have low
oscillator strengths, it is impossible to observe them in absorption. In some
isoelectronic sequences, close above the ground state of the ion there exist
metastable levels that can have a significant population. A well known example
is the Be-sequence, where the \ion{C}{iii} line at 1909~\AA\ is often used as a
density diagnostic in plasmas. Another ion in this iso-electronic sequence is
\ion{O}{v}, where the metastable 1s$^2$2s\,2p~$^3$P level is only 10~eV above
the ground state 1s$^2$2s$^2$~$^1$S$_0$. The $J=0$ term of this triplet cannot
decay radiatively to the ground state, leading to a significant population of
this level at all densities. K-shell absorption from the ground state of this
ion (at 22.3~\AA) has been observed in AGN (\object{NGC~5548}, Steenbrugge et
al. (\cite{steenbrugge03}, \cite{steenbrugge04}); \object{NGC~4051}, (Ogle et
al.\ \cite{ogle04}); and recently \object{Mrk~279} (Costantini et al.\
\cite{costantini04}). In this last source the spectrum indicates the possible
presence of absorption from the metastable level of \ion{O}{v}. The critical
density for the population of this metastable level is around
$10^{16}$~m$^{-3}$, a relevant density for AGN outflows. The presence or
absence of K-shell absorption lines from the metastable level of \ion{O}{v}
then is an important density diagnostic, which can serve to constrain distance
and geometry of the outflow.

In principle transitions from the metastable level of \ion{O}{v} can also be
studied using the UV absorption lines from the 2s\,2p triplet to 2p$^2$ triplet
near 760~\AA. Pettini \& Boksenberg (\cite{pettini86}) identified these lines
in IUE spectra of the BAL quasar \object{PG~0946+301}. However, in HST data of
the same object no evidence for these lines was found (Arav et al.
(\cite{arav99}). But in yet another BAL quasar (\object{QSO B0226-1024}) 
Korista et al. (\cite{korista92}) found evidence for the presence of these
lines, implying densities of the order of $10^{17}$~m$^{-3}$. However, due to
Galactic absorption such an analysis can only be done for bright, highly
redshifted quasars. For nearby Seyfert galaxies K-shell X-ray absorption lines
are the ideal tool to study densities.

In this paper we study the expected population of the metastable level of
\ion{O}{v} under photoionised conditions, as well as the wavelengths and
oscillator strengths of the corresponding K-shell absorption lines. We then
present a possible detection of absorption from \ion{O}{v*} levels in the
Chandra LETGS spectrum of Mrk~279.

\section{Level populations}

\subsection{Transition rates}

\begin{table}[!h]
\caption{Energy levels of \ion{O}{v}. Energies are taken from Wiese et al.\
(\cite{wiese96}). Levels 2--4 are the metastable levels discussed in this
paper.}
\label{tab:levels}
\centerline{
\begin{tabular}{rrrr}
\hline
\hline
level & Configuration & $J$ & Energy (eV) \\
\hline
1 & 2s$^2$~$^1$S    & 0 & 0 \\
2 & 2s\,2p~$^3$P    & 0 & 10.16 \\
3 &                 & 1 & 10.18 \\
4 &                 & 2 & 10.21 \\
5 & 2s\,2p~$^1$P    & 1 & 19.69 \\
6 & 2p$^2$~$^3$P  & 0 & 26.47 \\
7 &               & 1 & 26.49 \\
8 &               & 2 & 26.52 \\
9 & 2p$^2$~$^1$D  & 2 & 28.73 \\
10& 2p$^2$~$^1$S  & 0 & 35.70 \\
\hline\noalign{\smallskip}
\end{tabular}
}
\end{table}

\begin{figure}
\resizebox{\hsize}{!}{\includegraphics[angle=0]{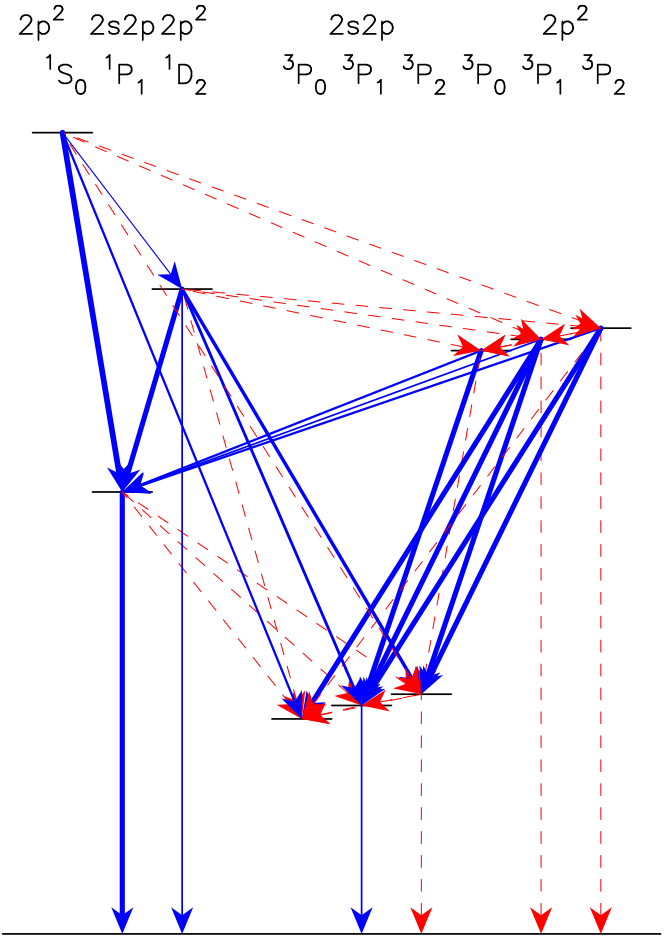}}
\caption{Energy level diagram of \ion{O}{v}. Only the $n=2$ levels are shown.
The energy differences within the $^3$P triplets have been exaggerated for
clarity. Solid lines: transitions with transition probabilities larger than
1\,s$^{-1}$; the line thickness is proportional to the logarithm of the
transition probability. Dashed lines: transitions with transition probabilities
less than  1\,s$^{-1}$.}
\label{fig:levels}
\end{figure}

We made a model for the population of all 10 levels with principal quantum
number $n=2$. These levels are listed in Table~\ref{tab:levels}. The
energy-level diagram is shown in Fig.~\ref{fig:levels}. We omit the $n>2$
levels from our calculation, as the lowest of these levels has an energy of
67.82~eV. This is too high as compared to the typical temperatures of a few eV
for which \ion{O}{v} is formed under photoionised conditions.

The population levels $n_i$ were solved using the equation
\begin{equation}
R_{ij}n_j = b_i,
\end{equation}
where we sum over $j$ and $R_{ij}$ is the total transition rate matrix from
level $j$ to level $i$, and $b_i$ represents source/loss terms like ionisation
from \ion{O}{iv} to \ion{O}{v} or recombination from \ion{O}{vi} to \ion{O}{v}.
Other loss terms like ionisation or recombination of \ion{O}{v} itself can be
accomodated in $R_{ij}$, as these processes are proportional to the \ion{O}{v}
density.

Radiative transition rates between the levels of Table~\ref{tab:levels} were
taken from the compilation of Wiese et al. (\cite{wiese96}).

We took the excitation rates from Safranova et al.\ (\cite{safranova95}) for
the following transitions: 2s$^2$--2s\,2p and 2s\,2p--2p$^2$. The remaining
rates were taken from Kato et al.\ (\cite{kato90}). Kato et al.\ give only the
total rates for the following transitions involving a singlet and a triplet:
2s$^2$--2p$^2$~$^3$P, 2s\,2p~$^3$P--2s\,2p~$^1$P, 2p$^2$~$^3$P--2p$^2$~$^1$D
and 2p$^2$~$^3$P--2p$^2$~$^1$S. Lacking more information, we therefore
subdivided these rates according to the statistical weights of the triplets
involved. This is also found to be valid for the distorted wave calculations of
Zhang \& Sampson (\cite{zhang92}), which explicitly give multiplet-resolved
rates. There is a caveat, however. As noted by Safranova et al.\
(\cite{safranova95}) in the forbidden transitions ($J^\prime-J = 0-0, 0-2,
2-0$), only the exchange contribution to the collision strength is important,
and the direct part is zero. The $^3$P$_0 - ^1$D$_2$, $^3$P$_0 - ^1$S$_0$ and
$^3$P$_2 - ^1$S$_0$ transitions of the above mentioned configurations are
forbidden and would suffer from this effect. As however in most astrophysical
situations the occupation of the $2p^2$ levels is low, and these levels have
strong radiative transitions to lower levels, the errors made by our
approximation are not too serious.

We added the proton excitation rates within the 2s\,2p and 2p$^2$ $^3$P
triplets from Doyle et al.\ (\cite{doyle80}).

We also included radiative recombination (\ion{O}{vi} to \ion{O}{v} and
\ion{O}{v} to \ion{O}{iv}) and collisional ionisation rates (\ion{O}{iv} to
\ion{O}{v} and \ion{O}{v} to \ion{O}{vi}). We took the total radiative
recombination rates from Arnaud \& Rothenflug (\cite{arnaud85}) and assume that
the rates towards the $n=2$ levels, either direct or indirect, are proportional
to the statistical weight of these levels. The collisional ionisation rates
were also taken from Arnaud \& Rothenflug (\cite{arnaud85}). For ionisation of
\ion{O}{iv}, we assume that all \ion{O}{iv} ions are in the 2s$^2$2p~$^2$P
ground state, and that 2s-shell ionisation of \ion{O}{iv} results into a 
2s\,2p~$^3$P state of \ion{O}{v} with the population of the sublevels
distributed according to their statistical weight. Note however that for the
parameter range that we consider the collisional ionisation and radiative
recombination rates are relatively unimportant.

At low temperatures charge transfer reactions with neutral hydrogen become
important. We included the charge transfer recombination rates for
recombination of \ion{O}{vi} and \ion{O}{v} from Rakovic et al.\
(\cite{rakovic01}). Similar to the radiative recombination, we assume that the
rates towards the $n=2$ levels, either direct or indirect, are proportional to
the statistical weight of these levels.

\subsection{Collisional ionisation equilibrium}

Using the rates determined in the previous section, we determine equilibrium
population levels for a range of densities and temperatures, first under
collisional ionisation equilibrium (CIE) conditions. We restrict the
temperature range to $kT<20$~eV; higher temperatures are not expected under
photoionised conditions, and moreover they would result in a significant amount
of excitation to the $n=3$ levels, with the corresponding downwards radiative
decays. 

For this temperature range, the effect of collisional ionisation is negligible,
as the electrons have insufficient energy to overcome the 113.896~eV ionisation
potential.  Radiative recombination affects the population of the 2s\,2p~$^3$P
triplet by no more than 3~\%. Proton excitation modifies the populations only
significantly (more than 3~\%) for temperatures higher than 10~eV, but even for
$kT=30$~eV the effects are smaller than 11~\% for all densities.

Charge transfer recombination due to collisions with \ion{H}{i} atoms appears
to be very important at temperatures below 2~eV. This is mainly due to the
large recombination cross section at low temperatures and the high neutral
fraction at these low temperatures. Since under CIE conditions the \ion{O}{v}
fraction is less than 0.001 for temperatures below 9~eV, inclusion of charge
transfer recombination is of little practical use under those circumstances,
however.

\begin{figure}
\resizebox{\hsize}{!}{\includegraphics[angle=-90]{cierat.cps}}
\caption{Population of the 2s\,2p~$^3$P triplet with respect to the
ground state in \ion{O}{v} as a function of density and $kT$ (in eV) as
labeled. All three sublevels ($J=0$, $J=1$ and $J=2$) have been added.}
\label{fig:cierat}
\end{figure}

Fig.~\ref{fig:cierat} shows the relative occupation $P$ of the 2s\,2p~$^3$P
triplet with respect to the ground state. The enhancement of this ratio at
$kT=0.5$~eV and $kT=1$~eV is due to the charge transfer recombinations. The
sharp increases in $P$ as a function of hydrogen density $n_{\mathrm H}$ around
densities of a) $10^{12}$, b) $10^{17}$ and c) $10^{22}$~m$^{-3}$ are due to
the following. At a), the collisional coupling of the 2s\,2p~$^3$P$_2$ level
with the 2s\,2p~$^3$P$_1$ level (either direct or via excitation to the
2p$^2$~$^3$P$_1$ level followed by radiative decay) becomes stronger than the
radiative decay by a forbidden transition to the ground state. At b), the 
collisional coupling of the 2s\,2p~$^3$P$_1$ level with the other triplet
levels is stronger than its radiative decay to the ground state. Finally, at c)
the ground state depopulates due to strong collisional coupling of the
2s\,2p~$^1$P level with the 2p$^2$~$^1$D,$^1$S levels. This causes the ratio of
the triplet with respect to the ground state to increase (the ratio with
respect to the sum of all \ion{O}{v} levels does not change).

\subsection{Photoionisation equilibrium\label{sect:pie}}

For photoionised plasmas we use the same approach as before, only the
transition matrix $R_{ij}$ and the source term $b_i$ contains more terms. We
now include both photo-excitation and stimulated emission between the $n=2$
levels. As we will use an ionising spectrum with a strong UV component (see
below), photo-excitations towards higher levels than $n=2$ are relatively
unimportant. We include photoionisation of \ion{O}{iv} from both the 2p and 2s
shells. For our ionising spectrum, the photoionisation rate of \ion{O}{iv} from
the 1s shell is only 2.5~\% of the 2p rate, and as 1s ionisation of \ion{O}{iv}
is followed in most cases by an Auger transition, producing \ion{O}{vi}, we
neglect this process here. We also neglect the similar process of K-shell
ionisation of \ion{O}{iii} followed by an Auger transition, as we assume that
the concentration of \ion{O}{iii} relative to \ion{O}{v} may be small and again
K-shell photoionisation is rare relative to L-shell photoionisation for our
ionising spectrum. However, we included the 5~\% contribution of 1s
photoionisation of \ion{O}{v} to the total $n=2$ photoionisation rate, as this
represents a loss term for which we are not interested in the end product.

\begin{figure}
\resizebox{\hsize}{!}{\includegraphics[angle=-90]{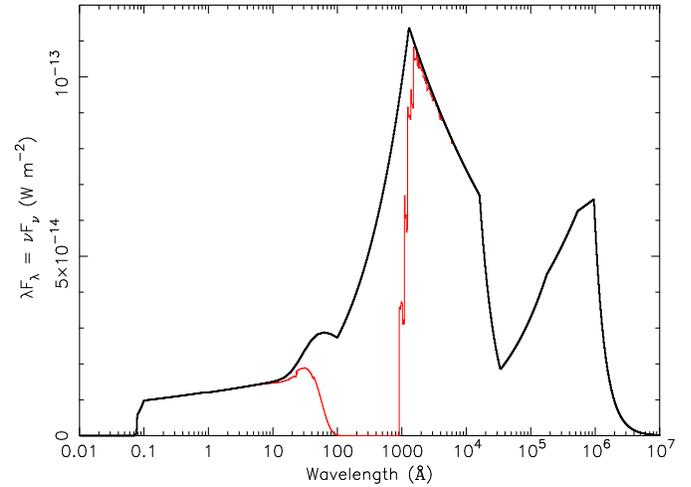}}
\caption{Spectral energy distribution (thick line)
used for our photoionisation
calculations. The thin line shows the same spectrum absorbed by a Galactic
foreground of cold gas with a hydrogen column density of $1.64\times
10^{24}$~m$^{-2}$ and $\sigma_v = 10$~km\,s$^{-1}$.}
\label{fig:sed}
\end{figure}

\begin{figure}
\resizebox{\hsize}{!}{\includegraphics[angle=-90]{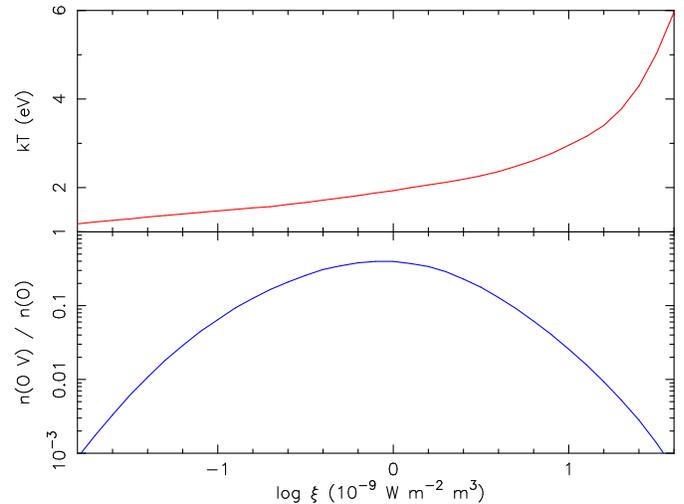}}
\caption{Temperature (upper panel) and \ion{O}{v} fraction (lower panel) as a
function of ionisation parameter $\xi$. The maximum \ion{O}{v} concentration of
40~\% occurs for $\log \xi = -0.05$, at a temperature $kT=1.9$~eV (to be
compared with 47~\% at $kT=21.0$~eV under CIE conditions).}
\label{fig:xit_cloudy}
\end{figure}

For the photoionising spectrum we adopt a model based upon the observed 
Chandra and UV continuum of Mrk~279 (Costantini et al. \cite{costantini04}). 
This model is similar to the model employed for NGC~5548 by Kaastra et al.
(\cite{kaastra02}). The spectrum contains a strong UV bump
(Fig.~\ref{fig:sed}). We normalized it to a 1--1000~Ryd luminosity $L$ of
$8.31\times 10^{37}$~W, in order to reproduce the observed UV and X-ray
continuum of Mrk~279. Note that for this spectral shape the relation between
the photoionisation parameters $\xi$ and $U$ is given by $U=Q/4\pi
r^2nc=0.036\xi$, where $Q$ is the number of hydrogen ionising photons and $\xi$
is expressed in units of $10^{-9}$~W\,m. The concentrations of
\ion{O}{iv}--\ion{O}{vi} as well as the temperature $T(\xi)$ as a function of
the ionisation parameter $\xi$ were obtained from a set of runs with Cloudy
Version 95.06 (Ferland \cite{ferland02}) for a thin photoionised slab at low
density and with solar abundances. In that approximation, $T(\xi)$ and the ion
concentrations only depend upon the shape of the ionizing spectrum. 
Fig.~\ref{fig:xit_cloudy} shows the \ion{O}{v} concentration and $T(\xi)$
relation. We then determined the level populations as a function of $n_{\mathrm
H}$ and $T$; inverting $T(\xi)$ yields $\xi$ and using $\xi$, $L$ and
$n_{\mathrm H}$ we then know the distance of the absorber to the central source
and hence the relevant fluxes for photoionisation and photo-excitation. 

\begin{figure}
\resizebox{\hsize}{!}{\includegraphics[angle=-90]{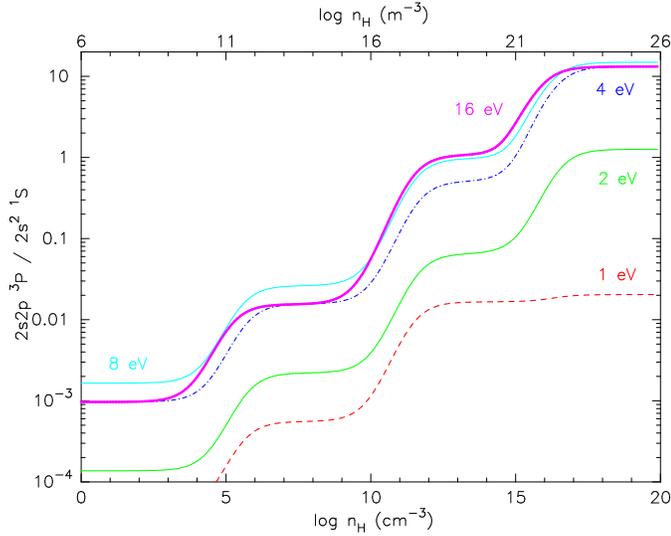}}
\caption{Population of the 2s\,2p~$^3P$ triplet with respect to the
ground state in \ion{O}{v} as a function of density and $kT$ (in eV) as
labeled for a photoionised plasma. 
All three sublevels ($J=0$, $J=1$ and $J=2$) have been added.}
\label{fig:pierat}
\end{figure}

We show our results in Fig.~\ref{fig:pierat}. At low temperatures ($kT < 1$~eV)
the neutral hydrogen fraction is now much smaller than in a collisional
equilibrium plasma, resulting in less important effects of charge transfer
recombination. For temperatures $kT$ between 2--4~eV and densities below
$10^{20}$~m$^{-3}$ the results for photoionisation equilibrium (PIE) and CIE
are not significantly different. At higher densities photo-excitation of in
particular the 2s$^2$--2s\,2p~$^1$P$_1$ resonance line (629.73~\AA) is more
important than collisional excitation from the  ground state to the
2s\,2p~$^1$P$_1$ level, causing enhanced higher level populations.

\subsection{Resonance line trapping}

In Sect.~\ref{sect:pie}, our treatment of PIE assumed that \ion{O}{v} is
located in a photoionised slab with negligible column density. However
modeling of the Mrk~279 spectrum shows that \ion{O}{v} has a column density of
order $10^{20}$~m$^{-2}$. For a velocity broadening $\sigma_v$ of
50~km\,s$^{-1}$ this implies that the 2s$^2$--2s2sp~$^1$P$_1$ resonance line
has an optical depth of 69 at the line center. As for a broad range of
parameter space radiative decay through this emission line is the dominant
decay proces of the upper level $j$, line photons effectively undergo resonance
line scattering. The direct escape probability $P_{\mathrm{esc}}$ per emitted
photon is therefore very small, and this leads effectively to a smaller
radiative decay rate $A_{ij}^\prime = P_{\mathrm{esc}} A_{ij}$, with $A_{ij}$
the true transition probability. We follow here the approach of  Hollenbach \&
McKee (\cite{hollenbach79}) and Kallman \& McCray (\cite{kallman82}) in order
to approximate $P_{\mathrm{esc}}$ for a slab:
\begin{equation} P_{\mathrm{esc}} = {(A_{ij}
/ A_{\mathrm{tot}}) \over 1. + \tau_0\sqrt{2\pi \ln(2.13+\tau_0^2)}},
\end{equation} 
where $\tau_0$ is the optical depth at line center, and $A_{\mathrm{tot}}$ is
the total transition probability from level $j$ to any higher or lower level,
including radiative and collisional processes. Resonance line trapping appears
to be relevant for the transition between levels 1--5 (629.73~\AA) mentioned
above, as well as for the 3--6 (761.13~\AA), 5--9 (1371.30~\AA) and 5--10
(774.52~\AA) transitions whenever the 2s\,2p configuration has a significant
population.

\begin{figure}
\resizebox{\hsize}{!}{\includegraphics[angle=-90]{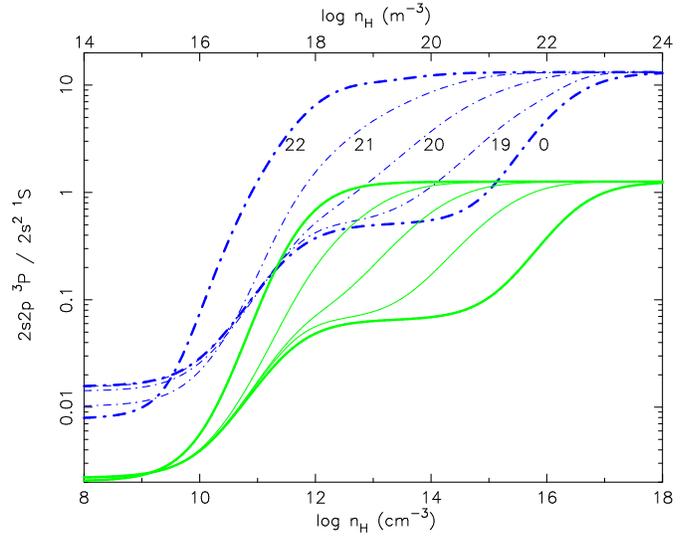}}
\caption{Population of the 2s\,2p~$^3P$ triplet with respect to the
ground state in \ion{O}{v} as a function of density and $kT$ (in eV) in a
photoionised plasma. Solid curves are for $kT=2$~eV, dash-dotted curves
for $kT=4$~eV. Labels indicate $\log N$ with $N$ the column density
of \ion{O}{v} in m$^{-2}$. 
All three sublevels ($J=0$, $J=1$ and $J=2$) have been added.}
\label{fig:trap}
\end{figure}

In Fig.~\ref{fig:trap} we show the results for our calculations including
resonance line trapping. It is seen that the critical density at which the
2s\,2p triplet level gets a significant population can shift downwards by
several orders of magnitude for sufficiently large column densities.

\section{X-ray absorption lines}

We have calculated wavelengths and oscillator strengths for all transitions
between levels with a K-shell vacancy (1s\,2s$^2$2p, 1s\,2s\,2p$^2$ and
1s2p$^3$) and the 1s$^2$\,2s$^2$, 1s$^2$\,2s\,2p, 1s$^2$\,2p$^2$ and
1s$^2$\,2s\,np levels. We used the code of Cowan (\cite{cowan81}) for this
purpose.

\begin{table}[!h]
\caption{Strongest X-ray lines of \ion{O}{v}. The values listed apply for
the full multiplets (not splitted into sublevels).}
\label{tab:lines}
\centerline{
\begin{tabular}{lllllcc}
\hline
\hline
label &
\multicolumn{2}{c}{Initial state} & 
\multicolumn{2}{c}{Final state} & 
$f$ & $\lambda$  \\
    & conf & term & conf     & term &       & (\AA) \\
\hline
A2 & 2s$^2$ & $^1$S & 1s\,2s$^2$2p   &        $^1$P & 0.649  & 22.381 \\
A3 & 2s$^2$ & $^1$S & 1s\,2s$^2$3p   &        $^1$P & 0.108  & 19.871 \\
A4 & 2s$^2$ & $^1$S & 1s\,2s$^2$4p   &        $^1$P & 0.041  & 19.258 \\
A5 & 2s$^2$ & $^1$S & 1s\,2s$^2$5p   &        $^1$P & 0.020  & 19.002 \\
A6 & 2s$^2$ & $^1$S & 1s\,2s$^2$6p   &        $^1$P & 0.012  & 18.869 \\
A7 & 2s$^2$ & $^1$S & 1s\,2s$^2$7p   &        $^1$P & 0.007  & 18.791 \\
A8 & 2s$^2$ & $^1$S & 1s\,2s$^2$8p   &        $^1$P & 0.005  & 18.742 \\
A9 & 2s$^2$ & $^1$S & 1s\,2s$^2$9p   &        $^1$P & 0.004  & 18.708 \\
B1 & 2s\,2p & $^3$P & 1s\,2s\,2p$^2$ & ($^3$S)$^3$P & 0.328  & 22.466 \\
B2 & 2s\,2p & $^3$P & 1s\,2s\,2p$^2$ & ($^3$S)$^3$D & 0.178  & 22.431 \\
B3 & 2s\,2p & $^3$P & 1s\,2s\,2p$^2$ & ($^3$S)$^3$S & 0.039  & 22.239 \\
B4 & 2s\,2p & $^3$P & 1s\,2s\,2p$^2$ & ($^1$S)$^3$P & 0.014  & 22.132 \\
C1 & 2s\,2p & $^1$P & 1s\,2s\,2p$^2$ & ($^1$S)$^1$D & 0.182  & 22.558 \\
C2 & 2s\,2p & $^1$P & 1s\,2s\,2p$^2$ & ($^3$S)$^1$P & 0.339  & 22.389 \\
C3 & 2s\,2p & $^1$P & 1s\,2s\,2p$^2$ & ($^1$S)$^1$S & 0.040  & 22.363 \\
D1 & 2p$^2$ & $^3$P & 1s\,2s$^2$\,2p &        $^3$P & 0.005  & 23.643 \\
D2 & 2p$^2$ & $^3$P & 1s\,2p$^3$     &        $^3$D & 0.173  & 22.511 \\
D3 & 2p$^2$ & $^3$P & 1s\,2p$^3$     &        $^3$S & 0.151  & 22.474 \\
D4 & 2p$^2$ & $^3$P & 1s\,2p$^3$     &        $^3$P & 0.105  & 22.339 \\
E1 & 2p$^2$ & $^1$D & 1s\,2s$^2$\,2p &        $^1$P & 0.006  & 23.598 \\
E2 & 2p$^2$ & $^1$D & 1s\,2p$^3$     &        $^1$D & 0.328  & 22.477 \\
E3 & 2p$^2$ & $^1$D & 1s\,2p$^3$     &        $^3$P & 0.002  & 22.457 \\
E4 & 2p$^2$ & $^1$D & 1s\,2p$^3$     &        $^1$P & 0.112  & 22.306 \\
F1 & 2p$^2$ & $^1$S & 1s\,2s$^2$\,2p &        $^1$P & 0.002  & 23.855 \\ 
F2 & 2p$^2$ & $^1$S & 1s\,2p$^3$     &        $^1$P & 0.438  & 22.536 \\ 
\hline\noalign{\smallskip}
\end{tabular}
}
\end{table}

Table~\ref{tab:lines} lists the calculated wavelenghts and oscillator strengths
$f$ of all lines with $f>0.001$ and principal quantum number $n<10$. We
determined only total oscillator strengths and average wavelengths of the
multiplets, as the current X-ray instrumentation has insufficient resolution to
resolve the individual lines of the multiplets. 

An important question is the accuracy of the wavelengths. In general these
wavelengths have never been measured in the laboratory, with the exception of
the resonance transition A2. This wavelength is 22.374$\pm$0.003~\AA\ (Schmidt
et al.\ \cite{schmidt04}).

The lower levels (1s$^2$\,2$\ell$\,$2\ell^\prime$) of these transitions have
accurately known energy levels $E_i$  (Wiese et al.\ \cite{wiese96}). For the
upper levels (1s\,2$\ell$\,2$\ell^\prime$\,n$\ell^{\prime\prime}$) accurate
energy levels $E_j$ can be obtained indirectly from Auger electron spectra
through the relation $E_j = E_{\mathrm A} + I$, with $I$ the ionisation
potential of \ion{O}{v}, for which we take 113.896~eV (Wiese et al.\
\cite{wiese96}), and $E_{\mathrm A}$ the energy of the Auger electron. 

\begin{table}[!t]
\caption{Core-excited energy levels in \ion{O}{v}. Columns (1) and (2):
configuration and term; (3) and (4): Calculated energy of the Auger electron
$E_{\mathrm A}$ for a transition to the 1s$^2$\,2s~$^2S$ (ground) state of
\ion{O}{vi}; where only transition energies to the 1s$^2$\,2p~$^2P$ state are
given in the original reference these have been corrected for the energy
difference between the $^2$S and $^2$P state; (5): measured $E_{\mathrm A}$;
(6): adopted $E_{\mathrm A}$; (7): adopted energy $E_j=E_{\mathrm A}+I$ of the
level.}
\label{tab:auger}
\centerline{
\begin{tabular}{@{\,}l@{\,\,}l@{\,\,}l@{\,\,}l@{\,\,}l@{\,\,}l@{\,\,}l@{\,}}
\hline
\hline
conf. & term & 
$E_{\mathrm A}$ (th) &
$E_{\mathrm A}$ (th) &
$E_{\mathrm A}$ (exp) &
$E_{\mathrm A}$  &
$E_j$ \\
 & & (eV) & (eV) & (eV) & (eV) & (eV) \\
\hline
1s\,2s$^2$\,2p & $^1$P       & 441.16$^a$& 440.55$^d$& 440.5$\pm$0.2$^f$  & 440.55 & 554.45$\pm$0.20\\
               & $^3$P       & 436.63$^a$& 435.96$^b$& 435.9$\pm$0.2$^f$  & 435.96 & 549.86$\pm$0.20\\
1s\,2s$^2$\,3p & $^1$P       &           & 507.54$^d$&  		  & 507.54 & 621.44$\pm$0.20\\
1s\,2s\,2p$^2$ & $^1$S       & 462.10$^a$& 460.83$^c$& 460$\pm$2$^e$      & 460.83 & 574.73$\pm$0.20\\
               & $^1$P       & 461.98$^a$& 460.12$^c$& 460.5$\pm$1$^e$    & 460.12 & 574.02$\pm$0.20\\
               & $^1$D       & 456.74$^a$& 455.35$^c$& 456.3$\pm$0.8$^e$  & 455.35 & 569.25$\pm$0.20\\
               & $^3$S       & 455.06$^a$& 454.17$^b$& 454$\pm$1$^e$      & 454.17 & 568.07$\pm$0.20\\
               & ($^3$S)$^3$P& 448.76$^a$& 448.70$^b$& 448.6$\pm$0.2$^f$  & 448.70 & 562.60$\pm$0.20\\
               & ($^1$S)$^3$P& 458.58$^a$& 456.9$^e$ &  		  & 457.50 & 571.40$\pm$0.60\\
               & $^3$D       & 449.74$^a$& 448.68$^b$& 448.5$\pm$1$^e$    & 448.68 & 562.58$\pm$0.20\\
               & $^5$P       & 438.33$^a$&	     & 438.55$\pm$0.2$^f$ & 438.55 & 552.45$\pm$0.20\\
1s\,2p$^3$     & $^1$P       & 473.72$^a$& 471.80$^d$&  		  & 471.80 & 585.70$\pm$0.20\\
               & $^1$D       & 468.05$^a$&	     &  		  & 466.97 & 580.87$\pm$0.60\\
               & $^3$S       & 464.91$^a$&	     &  		  & 463.83 & 577.73$\pm$0.60\\
               & $^3$P       & 469.40$^a$&	     &  		  & 468.32 & 582.22$\pm$0.60\\
               & $^3$D       & 463.75$^a$&	     &  		  & 462.67 & 576.57$\pm$0.60\\
               & $^5$S       & 456.27$^a$&	     & 456.4$\pm$0.2$^f$  & 456.40 & 570.30$\pm$0.20\\
\hline\noalign{\smallskip}
\end{tabular}
}
\begin{list}{}{}
\item[$^{\mathrm{a}}$] Chen \cite{chen85}
\item[$^{\mathrm{b}}$] Lin et al.\ \cite{lin01}
\item[$^{\mathrm{c}}$] Shiu et al.\ \cite{shiu01}
\item[$^{\mathrm{d}}$] Lin et al.\ \cite{lin02}
\item[$^{\mathrm{e}}$] Bruch et al.\ \cite{bruch79}
\item[$^{\mathrm{f}}$] Bruch et al.\ \cite{bruch87}
\end{list}
\end{table}

$E_{\mathrm A}$ has been measured in the laboratory or calculated
theoretically; we list the relevant values in Table~\ref{tab:auger}. A
comparison of the calculations of the group of K.T. Chung (Lin et al.\
\cite{lin01}; Shiu et al.\ \cite{shiu01}; Lin et al.\ \cite{lin02}) with the
measurements of Bruch et al.\ (\cite{bruch79}, \cite{bruch87}) shows excellent
agreement; the average difference between observed and calculated values is
$-$0.05$\pm$0.16~eV; we therefore adopt a nominal uncertainty of 0.20~eV for
these calculated energies, equal to the measurement errors in the best
measurements. A comparison of the energies calculated by Chen (\cite{chen85})
with the above mentioned calculations by the group of K.T. Chung shows an
average difference of 1.08$\pm$0.28~eV, with a standard deviation of 0.60~eV
for the 9 levels in common. Therefore whenever we use the energies calculated
by Chen, we subtract 1.08~eV and assign a nominal accuracy of 0.60~eV.

The adopted energies $E_j$ of Table~\ref{tab:auger} were combined with the 
lower level energies $E_i$ of Table~\ref{tab:levels} in order to calculate the
predicted wavelengths of the transitions from Table~\ref{tab:lines}. We list
those wavelengths in Table~\ref{tab:wavcomp}. Champeaux et al.
(\cite{champeaux03}) recently deduced the ionisation potential $I$ for
\ion{O}{v} to be 113.66$\pm$0.13~eV, based upon the quantum defect estimated
from resonances in the measured photoionisation cross section. If this value is
used instead of the value of 113.896~eV given by Wiese et al.\
(\cite{wiese96}), most wavelengths listed in the column "Auger" of
Table~\ref{tab:wavcomp} increase by 0.009~\AA.

A comparison of the energies of the 1s$^2$2$\ell$2$\ell^\prime$ levels
calculated by us using the Cowan (\cite{cowan81}) code with the values of
Table~\ref{tab:levels} (Wiese et al.\ \cite{wiese96}) shows that our calculated
energies are smaller by 0.71 (levels 2--4), 0.91 (level 5), 0.83 (levels 6--8),
0.16 (level 9) and 1.46~eV (level 10), respectively. Correcting all wavelengths
where these levels are involved by these differences and also adjusting all
K-shell vacancy states by the same number so as to get the laboratory
wavelength of the 22.374~\AA\ line correct, yields the column labeled with
"corr." in Table~\ref{tab:wavcomp}. The difference in wavelength between both
columns can be as large as 0.053~\AA, comparable to the energy resolution of
Chandra's LETGS. Finally, we list in Table~\ref{tab:wavcomp} also wavelengths
from the paper by Chen \& Crasemann (\cite{chen87}) as well as wavelengths
presently calculated with the Hebrew University Lawrence Livermore Atomic Code
(HULLAC).

\begin{table}[!h]
\caption{Wavelength comparison of the X-ray lines of \ion{O}{v}.
Wavelengths are given in \AA. The column "Cowan" gives $\lambda$ from
Table~\ref{tab:lines}; "corr." are these calculated wavelengths corrected
as described in the text; "Auger" are the wavelengths determined from
Auger energies as listed in Table~\ref{tab:auger}; "CC" are the calculations
by Chen \& Crasemann (\cite{chen87}); "HULLAC" are the wavelengths
calculated using the HULLAC code.}
\label{tab:wavcomp}
\centerline{
\begin{tabular}{llllll}
\hline
\hline
line & Cowan & corr.   & Auger            & CC     & HULLAC \\
\hline
A2 &  22.381  & 22.374 & 22.362$\pm$ 0.008 & 22.415 & 22.372 \\
A3 &  19.871  & 19.866 & 19.951$\pm$ 0.008 &        & 19.921 \\
A4 &  19.258  & 19.253 & 	           &        & 19.316 \\
A5 &  19.002  & 18.997 &	           &        & 19.060 \\
A6 &  18.869  & 18.864 &	           &        & 18.949 \\
A7 &  18.791  & 18.786 &	           &        & 18.849 \\
A8 &  18.742  & 18.737 &	           &        & 18.800 \\
A9 &  18.708  & 18.703 &	           &        & 18.766 \\
B1 &  22.466  & 22.488 & 22.444$\pm$ 0.008 & 22.535 & 22.425 \\
B2 &  22.431  & 22.453 & 22.445$\pm$ 0.008 & 22.494 & 22.415 \\
B3 &  22.239  & 22.260 & 22.224$\pm$ 0.008 & 22.279 & 22.197 \\
B4 &  22.132  & 22.153 & 22.092$\pm$ 0.024 & 22.139 & 22.091 \\
C1 &  22.558  & 22.589 & 22.561$\pm$ 0.008 & 22.668 & 22.522 \\
C2 &  22.389  & 22.419 & 22.367$\pm$ 0.008 & 22.453 & 22.365 \\
C3 &  22.363  & 22.393 & 22.338$\pm$ 0.008 & 22.448 & 22.349 \\
D1 &  23.643  & 23.670 & 23.690$\pm$ 0.009 & 23.790 & 23.572 \\
D2 &  22.511  & 22.538 & 22.540$\pm$ 0.024 & 22.615 & 22.478 \\
D3 &  22.474  & 22.501 & 22.492$\pm$ 0.024 & 22.566 & 22.404 \\
D4 &  22.339  & 22.365 & 22.311$\pm$ 0.024 & 22.383 & 22.331 \\
E1 &  23.598  & 23.597 & 23.584$\pm$ 0.009 & 23.747 & 23.520 \\
E2 &  22.477  & 22.476 & 22.455$\pm$ 0.024 & 22.584 & 22.430 \\
E3 &  22.457  & 22.456 & 22.400$\pm$ 0.024 & 22.529 & 22.479 \\
E4 &  22.306  & 22.305 & 22.260$\pm$ 0.008 & 22.353 & 22.283 \\
F1 &  23.855  & 23.908 & 23.901$\pm$ 0.009 & 24.112 & 23.765 \\
F2 &  22.536  & 22.589 & 22.543$\pm$ 0.008 & 22.677 & 22.503 \\
\hline\noalign{\smallskip}
\end{tabular}
}
\end{table}

In general, the theoretical calculations for the wavelengths differ quite a
lot. We illustrate this with the resonance line at 22.374~\AA.  Calculated
values range from 22.334~\AA\ (HULLAC) to 22.415~\AA\ (Chen \& Crasemann); for
comparison, Pradhan et al.\ (\cite{pradhan03}) give a value of 22.35~\AA.

\begin{table}[!h]
\caption{Oscillator strengths $f$, total transition 
probabilities $A_{\mathrm{rad}}$,
Auger decay rates $A_{\mathrm{A}}$ and total 
transition rates $A_{\mathrm{tot}}$ of \ion{O}{v}. Transition probabilities
are in units of $10^{12}$~s$^{-1}$.}.
\label{tab:trans}
\centerline{
\begin{tabular}{lllllrrr}
\hline
\hline
id & $f$\,$^a$ & $f$\,$^b$ & $A_{\mathrm{rad}}$\,$^a$ &
$A_{\mathrm{rad}}$\,$^b$ & $A_{\mathrm{A}}$\,$^{a,c}$ &
$A_{\mathrm{tot}}$\,$^d$ \\
\hline
  A2 &0.646  & 0.649  & 3.05  & 3.00  &87.9	 & 90.9\\
  A3 &       & 0.108  &       & 0.63  &129       &129.6\\
  A4 &       & 0.041  &       & 0.26  &129       &129.3\\
  A5 &       & 0.020  &       & 0.13  &129       &129.1\\
  A6 &       & 0.012  &       & 0.07  &129       &129.1\\
  A7 &       & 0.007  &       & 0.05  &129       &129.1\\
  A8 &       & 0.005  &       & 0.03  &129       &129.0\\
  A9 &       & 0.004  &       & 0.02  &129       &129.0\\ 
  B1 &0.326  & 0.328  & 4.29  & 4.34  &34.2	 & 38.5\\
  B2 &0.171  & 0.178  & 1.35  & 1.41  &94.2	 & 95.6\\
  B3 &0.039  & 0.039  & 1.56  & 1.56  &46.9	 & 48.5\\
  B4 &0.014  & 0.014  & 0.19  & 0.19  &100	 &100.2\\
  C1 &0.181  & 0.182  & 1.41  & 1.43  &213	 &214.4\\
  C2 &0.333  & 0.339  & 4.40  & 4.51  &22.9	 & 27.4\\
  C3 &0.041  & 0.040  & 1.63  & 1.60  &154	 &155.6\\
  D1 &0.007  & 0.005  & 0.08  & 0.06  &129	 &129.1\\
  D2 &0.169  & 0.173  & 1.32  & 1.37  &131	 &132.4\\
  D3 &0.153  & 0.151  & 6.00  & 5.99  &0.0015	 &  6.0\\
  D4 &0.102  & 0.105  & 1.36  & 1.42  &79.9	 & 81.3\\
  E1 &0.009  & 0.006  & 3.05  & 3.00  &87.9	 & 90.9\\
  E2 &0.330  & 0.328  & 4.31  & 4.35  &130	 &134.4\\
  E3 &0.000  & 0.002  & 1.36  & 1.42  &79.9	 & 81.3\\
  E4 &0.111  & 0.112  & 4.36  & 4.41  &76.9	 & 81.3\\
  F1 &0.003  & 0.002  & 3.05  & 3.00  &87.9	 & 90.9\\
  F2 &0.439  & 0.438  & 4.36  & 4.41  &76.9	 & 81.3\\
\hline\noalign{\smallskip}
\end{tabular}
}
\begin{list}{}{}
\item[$^{\mathrm{a}}$] Chen \& Crasemann \cite{chen87}
\item[$^{\mathrm{b}}$] Present work
\item[$^{\mathrm{c}}$] for A3--A9 see text
\item[$^{\mathrm{d}}$] Using $A_{\mathrm{tot}}=A_{\mathrm{rad}}+
A_{\mathrm{A}}$ with $A_{\mathrm{rad}}$ from the present work
\end{list}
\end{table}

There is a good agreement between our calculations for oscillator strengths and
transition probabilities and those of Chen \& Crasemann (\cite{chen87}), see
Table~\ref{tab:trans}. The rms difference between both sets of oscillator
strength is 0.003. Also, the total radiative transition probabilities from
core-excited levels agree within 2~\%. It is necessary to take also the Auger
transitions into account for estimating the lifetime of the core-excited
levels. We use the Auger rates determined by Chen \& Crasemann (\cite{chen87}).
Their rates are in good agreement with the calculations of Lin et al.\
(\cite{lin01}, \cite{lin02}) for the 6 levels in common. These Auger rates are
an order of magnitude larger than the radiative transition rates. Thus
neglecting Auger rates will affect the estimated line equivalent width for high
column densities. There are no Auger rates available for the A4--A9
transitions. Lin et al.\ (\cite{lin02}) calculate an Auger rate of $1.29\times
10^{14}$~s$^{-1}$ for the upper level of transition A3. In their calculations
for Be, they find that for higher principal  quantum number $n$ the Auger rates
do not depend strongly upon $n$. Therefore we adopt a constant value for
$A_{\mathrm{A}}$ for transitions A3--A9.

\section{Comparison with the observed spectrum of Mrk~279}

\subsection{Chandra LETGS observations}

\begin{figure}
\resizebox{\hsize}{!}{\includegraphics[angle=-90]{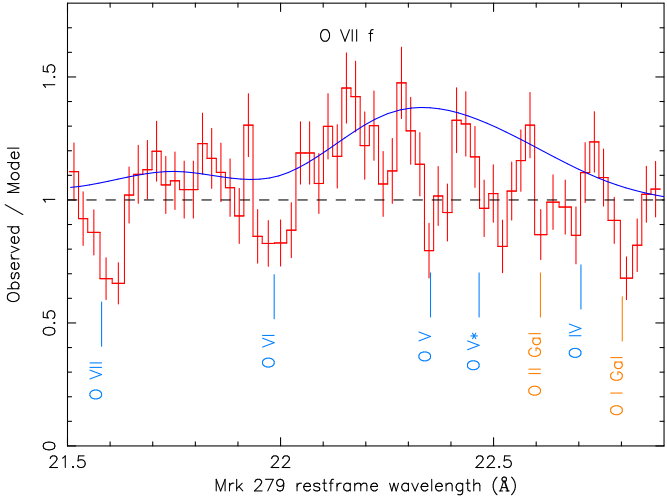}}
\caption{Chandra LETGS spectrum of Mrk~279 near the oxygen absorption lines.
The observed spectrum has been divided by the best fit simple continuum model
described in the text (dashed line). The additional broad emission feature
is indicated by the curved line. The wavelength scale is for the 
restframe of Mrk~279 (at $z=0.0306$). In the line identifications, we have
taken into account a small outflow (blueshift) of 300~km/s with respect
to this restframe; this outflow has been measured using simultaneous
high-resolution UV spectra. A few Galactic absorption lines are labeled with
"Gal".}
\label{fig:279ox}
\end{figure}

Mrk~279 is a moderately distant ($z=0.0306$) Seyfert~1 galaxy that we observed
for 340~ks with the Chandra LETGS. This instrument has a spectral resolution of
$\sim$0.05~\AA\ FWHM. The full data analysis of the LETGS spectrum of Mrk~279
will be presented elsewhere (Costantini et al.\ \cite{costantini04}). Here we
only summarize the results relevant for the study of the \ion{O}{v} absorption
lines. The LETGS spectrum was fit by a simple power law plus modified blackbody
with Galactic absorption. This smooth continuum spectrum produces an excellent
fit to the data except for the few regions where Mrk~279 has absorption lines.
The column densities in Mrk~279 are several times smaller than in other
well-studied AGN, and therefore most absorption lines and edges are very weak
except for the lines from the most abundant metal, which is oxygen. These lines
are predominantly found in the 18--24~\AA\ band.

Our continuum fit is excellent everywhere except for the region near the 
K-shell transitions of oxygen (Fig.~\ref{fig:279ox}). Here a broad emission
feature with a peak amplitude of about 30~\% of the continuum is visible.
Similar broad X-ray emission lines have been seen before in other AGN (Kaastra
et al.\ \cite{kaastra02}; Steenbrugge et al.\ \cite{steenbrugge04}), and may have
asymmetric line profiles. This broad line is due to the \ion{O}{vii} triplet
from the broad line region. The presence of this broad feature makes the
estimate of the underlying continuum for the narrow absorption lines less
certain. We estimated the broad line profile by drawing a spline through the
spectral regions that are known to be line-free.

From a detailed study of the \ion{O}{vi} UV and X-ray lines (Gabel et al.
\cite{gabel04b}; Costantini et al.\ \cite{costantini04}) we find that the
dominant absorber has an outflow velocity of 300~km/s and a line profile that
can be approximated to first order by a Gaussian with $\sigma_{\mathrm v} =
50$~km\,s$^{-1}$.

\subsection{Observed \ion{O}{v} lines}

We clearly see the absorption at 22.37~\AA, which we identify as the
1s$^2$\,2s$^2$  $-$ 1s\,2s$^2$\,2p~$^1$P$_1$ transition of \ion{O}{v}
(Fig.~\ref{fig:279ox}). Its measured wavelength in the rest frame of Mrk~279,
corrected for the 300~km/s outflow velocity, is 22.377$\pm$0.009~\AA, in
excellent agreement with the laboratory wavelength of 22.374~\AA\ mentioned
before (Schmidt et al. \cite{schmidt04}. The quoted error bar is only the
statistical error; we estimate the systematic uncertainty (due to calibration
uncertainties in the HRC-S detector) to be 0.008~\AA. 

\begin{table}[!h]
\caption{Observed absorption lines in Mrk~279, due to or related to \ion{O}{v}.
The observed wavelength $\lambda$ is given with two error bars: the first is
the statistical uncertainty, the second the systematic due to the wavelength
calibration uncertainty of the LETGS. For lines with wavelength fixed, the
observed equivalent width (EW) is merely the formal upper limit. The line
around 22.50~\AA\ is either fitted by a single line (22.540~\AA) or
alternatively by two lines (22.501 and 22.548~\AA). }
\label{tab:obslines}
\centerline{
\begin{tabular}{lllllrrr}
\hline
\hline
$\lambda$ (\AA) & EW (m\AA) & identification \\
\hline
22.377$\pm$0.009$\pm$0.008 & 18$\pm$6 & A2        \\
19.957 (fixed)             &  2$\pm$8 & A3        \\
19.325 (fixed)             &  6$\pm$6 & A4        \\      
22.540$\pm$0.009$\pm$0.010 & 20$\pm$5 & B1+B2?; see text \\
22.501$\pm$0.016$\pm$0.010 & 10$\pm$6 & B1+B2?; see text \\
22.548$\pm$0.010$\pm$0.010 & 18$\pm$6 & ?; see text \\
22.269$\pm$0.015$\pm$0.006 & 10$\pm$6 & B3? ; see text \\
\hline\noalign{\smallskip}
\end{tabular}
}
\end{table}

We find no clear evidence for transition A3. Upper limits to its equivalent
width depend upon the adopted wavelength. In another Seyfert~1 galaxy with a
strong A2 line we find evidence for a 2$\sigma$ detection of transition A3 at a
laboratory wavelength of 19.957$\pm$0.015~\AA, right within the wavelength
range indicated in Table~\ref{tab:wavcomp}. We adopt this wavelength for
transition A3 and use it to measure the limits to the line equivalent width
(Table~\ref{tab:obslines}). We also do not detect line A4. We take here the
wavelength as calculated with the HULLAC code to estimate the equivalent width.
Using the velocity broadening $\sigma_{\mathrm v}$ of 50~km/s and the
equivalent widths of lines A2, A3 and A4, we find a column density of
$10^{{20.16\pm0.42}}$~m$^{-2}$ for \ion{O}{v} in its ground state.
The large uncertainty is mainly due to the strong saturation of the A2 line, 
which has an optical depth of 3.7 at its line center.

Most absorption lines in the spectral band displayed in Fig.~\ref{fig:279ox}
can be easily identified by Galactic absorption lines or absorption lines due
to the warm absorber. Only the feature around 22.50~\AA\ has no obvious
identification with absorption from the ground state of any abundant ion,
either in the warm absorber or the Galactic foreground. When fitted by a single
line, its laboratory wavelength, corrected for the 300~km/s outflow, is 
22.540$\pm$0.009~\AA. A marginally better fit ($\Delta \chi^2 = -3.6$ for two
additional degrees of freedom) is obtained for a double, blended line, with
laboratory wavelengths of 22.501$\pm$0.016~\AA\ and 22.548$\pm$0.010~\AA,
respectively (see Table~\ref{tab:obslines}). Again, quoted errors are
statistical only; the systematic uncertainty is 0.010~\AA\ and is not
correlated with the systematic uncertainty in the main 22.374~\AA\ line. 

We suggest that in the case of a double blended line, the 22.501~\AA\ component
of the blend may be due to a combination of the B1 and B2 lines of \ion{O}{v},
which have a wavelength in the 22.47--22.53~\AA\ range. For the single line
solution (22.540~\AA) this is somewhat less likely. We note that the added
statistical and systematic uncertainty of the measurement of the line centroid
is in the 0.02--0.03~\AA\ range. The problem with a firm identification lies of
course in both the uncertainty in the theoretical wavelength as well as in the 
measured wavelength. 

\begin{figure}
\resizebox{\hsize}{!}{\includegraphics[angle=-90]{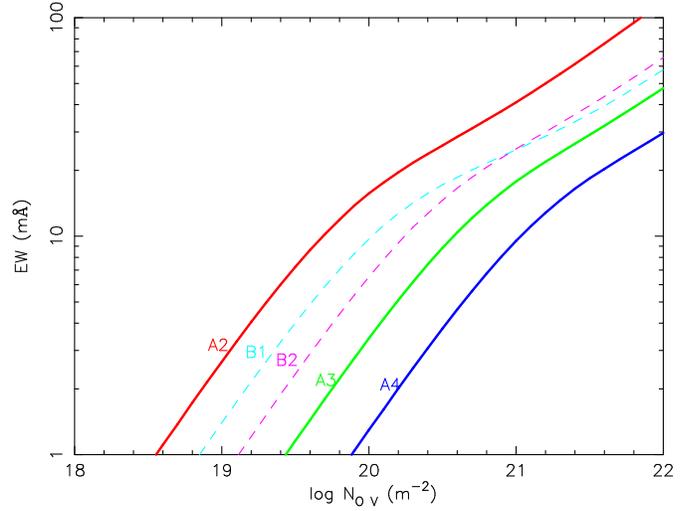}}
\caption{Curve of growth for the most important K-shell absorption lines
of \ion{O}{v}, for a Gaussian velocity broadening $\sigma_v$ 
of 50~km\,s$^{-1}$.}
\label{fig:ew}
\end{figure}

If our identification is true, we can deduce the column density of \ion{O}{v}
in its 2s\,2p~$^3$P triplet state. From the curve of growth (see
Fig.~\ref{fig:ew}) and an equivalent width of 10$\pm$6~m\AA, we find a column
of $10^{{19.82\pm0.44}}$~m$^{-2}$. This combined with the column density of
\ion{O}{v} in its ground state yields a value for the population of the
2s\,2p~$^3$P triplet state relative to the ground state of 0.5, with rms error
bounds of a factor of 4 (i.e, the relative population is between 0.125 and 2).
From the column density of the 2s\,2p~$^3$P triplet state we predict that line
B3 should have an equivalent width of 1~m\AA\, with an upper limit of 3~m\AA.
The feature at 22.269~\AA\ (see Table~\ref{tab:obslines}) could be marginally
consistent with this.

Table~\ref{tab:denslim} list the densities derived from the population ratio
for a few temperatures.

\begin{table}[!h]
\caption{Densities (in m$^{-3}$) for 2s\,2p triplet to ground ratio's of 0.125,
0.5 and 2, for a total \ion{O}{v} column density of $1.5\times
10^{20}$~m$^{-2}$.}
\label{tab:denslim}
\centerline{
\begin{tabular}{lllllrrr}
\hline
\hline
k$T$ (eV) & 0.125 & 0.5 & 2 \\
\hline
2 & $2.5\times 10^{18}$ & $3\times10^{19}$  & $\infty$ \\
4 & $10^{17}$           & $10^{18}$         & $1.5\times 10^{19}$ \\
6 & $5\times 10^{16}$   & $4\times 10^{17}$ & $10^{19}$ \\
\hline\noalign{\smallskip}
\end{tabular}
}
\end{table}

From our photoionisation equilibrium calculations, it follows that \ion{O}{v}
has its maximum concentration for k$T=2$~eV (see Fig.~\ref{fig:xit_cloudy}).
This yields a lower limit to the density of $2.5\times 10^{18}$~m$^{-3}$. 
k$T=2$~eV corresponds to $\xi = 0.9\times 10^{-9}$~W\,m$^{-2}$\,m$^{3}$, which
then leads to an upper limit to the distance $r$ of $2\times 10^{14}$~m or one
light week. This would place the \ion{O}{v} absorber close to the inner broad
line region. Although not completely impossible, such a small radius is not
very plausible, because Balmer line time-delay measurements suggest a broad
emission line region size of 6--17 light days for Mrk~279 (Scott et al.\
\cite{scott04} and references therein), and UV spectra show that the broad
lines are covered by the warm absorber.  

The temperature may be higher than 2~eV, however, if additional heating
processes occur in the absorbing region. Cloudy (Ferland \cite{ferland02})
allows for the inclusion of additional heat sources. A temperature of 6~eV is
reached for $\xi = 0.7\times 10^{-9}$~W\,m$^{-2}$\,m$^{3}$, with an additional
heating rate of  $1.0\times 10^{-7}n_{14}^2$~W\,m$^{-3}$, where $n_{14}$ the
density in units of $10^{14}$~m$^{-3}$. In this case the additional heat source
dominates the total heating rate. This is also the case for all temperatures
above 3~eV, where the additional heating contributes more than 80~\% to the
total heating rate. We find for k$T=6$~eV a lower limit to the density of
$5\times 10^{16}$~m$^{-3}$, corresponding to an upper limit to the distance of
$1.5\times 10^{15}$~m or 59 light days. This would place the absorber outside
the broad line region.

Is there a possible heating process that can produce the additional heat? For
the lower limit to the density of $5\times 10^{16}$~m$^{-3}$, the required
heating rate $Q$ is 0.025~W\,m$^{-3}$. One possibility would be classical heat
conduction from a neighbouring hot region. Using a hydrogen column density of
$8\times 10^{23}$~m$^{-2}$ (corresponding to a typical \ion{O}{v} column of
$1.5\times 10^{20}$~m$^{-2}$) and the lower limit of $5\times 10^{16}$~m$^{-3}$
for the density, we find an upper limit to the size $d$ of the absorber of
$1.6\times 10^7$~m.  For a heat conductivity $\kappa$ of $10^{-10}T_{\mathrm
K}^{5/2}$~W\,K$^{-1}$\,m$^{-1}$ with $T_{\mathrm K}$ the temperature in K, and
putting the heating rate $Q$ due to conduction equal to $\kappa T / d^2$, we
find that we need a temperature of 300~eV for the hot gas surrounding the
absorbing medium. Hotter gas in photoionisation equilibrium coexisting with
colder gas is indeed known to be possible. As the geometry of the absorbing
regions is still not fully understood, we leave this scenario as an interesting
possibility.

In our above analysis, we assumed that as far as \ion{O}{v} is concerned the
physical conditions in the absorber are such that \ion{O}{v} has its maximum
relative concentration. It is of course also possible to have a hotter absorber
with a smaller concentration of \ion{O}{v}. For instance, without invoking an
additional heat source we find that k$T=4$~eV if $\xi$ is 20 times larger than
we used before, and correspondingly the \ion{O}{v}/O ratio is 100 times smaller.
In that case most of the oxygen should be in the form of \ion{O}{vii}. Our
column densities for \ion{O}{vi} and \ion{O}{vii} (see Costantini et al.\
\cite{costantini04}) are not inconsistent with such a scenario. In this case, we
avoid the additional heating problem, but of course we still require a high
density $n>10^{17}$~m$^{-3}$ in order to get the high metastable population
fraction. In this case however the distance to the central source is only
$2\times 10^{14}$~m or 8 light days.

We do not expect blending due to lines from higher levels to be important for
all temperatures below 4~eV (the line series labeled with C--F of
Table~\ref{tab:lines}). This is because for those temperatures the population
of these higher levels is small.

Finally, if the metastable level has a significant population, also the UV
absorption lines from this level should be detectable in principle.
Unfortunately the strong, allowed UV absorption lines from the metastable level
are in the unobserved part of the UV spectrum (near 760~\AA). If the density
would be very high such that also the 2s\,2p~$^1$P$_1$ level gets a high
population, then the strongest UV absorption line should be the transition
between levels 5--9 of Table~\ref{tab:levels} at 1371.30~\AA. Simultaneous HST
observations of Mrk~279 (Gabel et al. \cite{gabel04b}) yield an upper limit to
the column density for level 5 deduced from the non-detection of this line of
$5\times 10^{16}$~m$^{-2}$. Combined with the allowed range for the ground
state column density ($10^{19.82\pm 0.44}$~m$^{-2}$ as quoted before) this
leads to an upper limit of $10^{-3}$ for the population of level 5 with respect
to the ground. The corresponding upper limit to the density of $5\times
10^{17}$~m$^{-3}$ is consistent with the lower limits derived from the  X-ray
transition from the metastable level. It should be noted, however, that the
X-ray derived density constraints depend on several modeling and observational
factors, as is evident from our present discussion. 

\section{Conclusions}

We have investigated in this paper the population of the metastable level of
\ion{O}{v} under photoionised conditions. For sufficiently high densities and a
temperature of just a few eV the metastable levels can have a significant
population. The relevant critical densities also depend strongly upon the
column density, due to resonance line trapping. The population of higher levels
than the 2s\,2p $^3$P triplet is always relatively small. If the metastable
level has a significant population, then we should observe K-shell absorption
lines from these levels. The wavelengths of these lines are not very accurately
known, however. This makes the identification of these lines in the X-ray
spectrum of the Seyfert 1 galaxy Mrk~279 somewhat uncertain. If present,
however, then the observed equivalent widths imply densities of order
$10^{17}$~m$^{-3}$ or more, corresponding to a distance to the central source
of the order of a light month.

\begin{acknowledgements} 
Unfortunately, Rolf Mewe died unexpectedly just a week
before the first draft of this paper was finished. His broad overview and long
standing record in the field of X-ray spectroscopy, as well as his friendship
will not be forgotten by us. The Space Research Organization of the Netherlands
is supported financially by NWO, the Netherlands Organization for Scientific
Research. N.A. aknowledges NASA grants HST-GO-9688 and Chandra-04700532.
E. B. was supported by the Yigal-Alon Fellowship and by ISF grant
\#28/03. 
\end{acknowledgements}

\end{document}